\documentclass[12pt,table]{article}

\usepackage{url}
\usepackage{bbm}
\usepackage{mathtools}
\usepackage{amssymb}
\usepackage{amsthm}
\usepackage{empheq}
\usepackage{latexsym}
\usepackage{enumitem}
\usepackage{eurosym}
\usepackage{dsfont}
\usepackage{appendix}
\usepackage{color} 
\usepackage{hyperref}
\usepackage{frcursive}
\usepackage[utf8]{inputenc}
\usepackage[T1]{fontenc}
\usepackage{geometry}
\usepackage{multirow}
\usepackage{lmodern}
\usepackage{anyfontsize}
\usepackage{pgfplots}
\usepackage{stmaryrd}
\usepackage{natbib}
\usepackage{lscape}
\usepackage{xcolor}
\usepackage{capt-of}
\usepackage{float}

\setcitestyle{numbers,open={[},close={]},citesep={,}}

\pgfplotsset{compat=1.14}

\definecolor{red}{rgb}{0.7,0.15,0.15}
\definecolor{green}{rgb}{0,0.5,0}
\definecolor{blue}{rgb}{0,0,0.7}
\hypersetup{colorlinks, linkcolor={red},citecolor={green}, urlcolor={blue}}
			
\makeatletter \@addtoreset{equation}{section}

\newtheorem{theorem}{Theorem}[section]

\newtheorem{remark}[theorem]{Remark}

\usepackage{setspace}
\spacing{1.1}

\title{Liquidity Stress Testing using Optimal Portfolio Liquidation\footnote{This work benefits from the financial support of the Chaires Analytics and Models for Regulation, Financial Risk and Finance and Sustainable Development. Bastien Baldacci gratefully acknowledge the financial support of the ERC Grant 679836 Staqamof. Bastien Baldacci and Iuliia Manziuk would like to thank Mathieu Rosenbaum (Ecole Polytechnique), Kaitong Hu (Squarepoint Capital) and Olivier Guéant (Université Paris 1 Panthéon-Sorbonne) for fruitful discussions. Mike Weber would like to thank Robert Almgren (Quantitative Brokers) and Jim Gatheral (Baruch College) for helpful conversations. The opinions expressed in this paper belong solely to the authors.}}
\author{\makebox[.9\textwidth]{ Mike {\sc Weber}}\\ BlueCove Limited \\ mweber@bluecove.com \and Iuliia {\sc Manziuk}\\CMAP, Ecole Polytechnique\\ iuliia.manziuk@polytechnique.edu \and Bastien {\sc Baldacci}\\CMAP, Ecole Polytechnique\\ bastien.baldacci@polytechnique.edu }
\begin{document}
\setlength{\textfloatsep}{1\baselineskip plus 0.2\baselineskip minus 0.2\baselineskip}
\setlength{\floatsep}{1\baselineskip plus 0.2\baselineskip minus 0.2\baselineskip}
\setlength{\intextsep}{1\baselineskip plus 0.2\baselineskip minus 0.2\baselineskip}

\maketitle

\begin{abstract}
We build an optimal portfolio liquidation model for OTC markets, aiming at minimizing the trading costs via the choice of the liquidation time. We work in the Locally Linear Order Book framework of \cite{toth2011anomalous} to obtain the market impact as a function of the traded volume. We find that the optimal terminal time for a linear execution of a small order is proportional to the square root of the ratio between the amount being bought or sold and the average daily volume. Numerical experiments on real market data illustrate the method on a portfolio of corporate bonds. 
\end{abstract}

\section{Introduction}

The European Securities and Markets Authority (ESMA) has set out guidance on liquidity stress testing supplementary to the existing requirements enshrined in the AIFMD and UCITS directives, with the ESMA guidelines coming into force on 30th September 2020. The core of the liquidity stress testing framework is a model that can be used to estimate liquidation times and costs in a reasonably realistic way for a portfolio of investments, including funds that can take short positions such as hedge funds. The main components of a liquidity stress testing framework are predefined stress tests, a market liquidity model that estimates liquidation cost and time, and a governance framework. This paper focuses on the model of market liquidity applied to optimal portfolio liquidation for corporate bonds. The model needs to produce liquidation times and costs depending on the market volatility, market daily volumes, and bid-ask spreads. The trader faces a trade-off between liquidating quickly, resulting in unfavourable price changes or liquidating slowly, thus incurring Profit and Loss (hereafter PnL) volatility. \\

Since the seminal work of Almgren and Chriss in \cite{almgren2001optimal}, a vast literature on optimal execution has emerged. In the initial Almgren-Chriss framework, a trader, allowed to send only market orders, aims to liquidate a large position in one or several assets and minimize the direct costs. Directs costs include both transaction costs and market impact, where the latter refers to the fact that, on average, a large order moves the price in the sense of the order's direction (price goes up for a buy order and conversely for a sell order). Their framework has numerous extensions, such as models incorporating an order-flow, introducing a price limiter, considering both limit and market orders, and taking into account an adverse selection mechanism (see \cite{cardaliaguet2018mean,cartea2015optimal,cartea2016incorporating,huang2015mean,jaimungal2014optimal}). \\

The vast majority of the optimal liquidation models are designed for electronic markets with a central limit order book where the trader or broker can send limit and market orders. Surprisingly, the issue of optimal liquidation on OTC markets, especially for fixed income products such as corporate bonds or credit default swaps (CDS), has been the subject of little interest in academic research. In these markets, the notion of market microstructure is radically different as there is no market impact in the usual sense because of liquidity fragmentation. There is no unique source of liquidity but several dealers who receive requests for quotes (hereafter RFQ) for a given size of security. Given the price proposed by the dealer, the client can accept the transaction or suggest a better price. This particular mechanism of OTC markets is prone to more opacity, which has significant consequences for the nature of the price impact of a trade. For example, in \cite{hendershott2015click}, the authors show a notable price impact due to information leakage, when the client requests prices from several dealers. In \cite{shachar2012exposing}, the authors provide evidence of ``hot potato'' trading: an initial client's request changes the hands of dealers several times, while it is gradually incorporated into the price. The very notion of liquidity is ambiguous in such markets, and so is the price impact of a trade.\\

One of the works studying the price impact of individual transactions in an OTC market is \cite{eisler2016price}, where the authors focus on credit indices. They find only a quantitative (but not qualitative) difference in terms of order flow and price impact compared to electronic markets. Moreover, the study argues in favor of the idea of latent liquidity in OTC markets. This concept of liquidity modeling, introduced in \cite{toth2011anomalous}, suggests that there is a latent volume which is not revealed in the observable order book because of the trading strategies of the market participants. \\

In this work, we adapt the concept of latent liquidity in order to capture portfolio liquidation costs as a function of liquidation time based on a relatively small number of market data inputs such as estimated average daily volume (ADV), volatility and bid-ask spread. In Section \ref{sec_llob}, we recall the framework of the latent order book model of \cite{toth2011anomalous} and derive the price impact equation, which allows one to estimate the volume available in a price range $[p,p+\Delta p]$, where $p$ is a certain price level and $\Delta p>0$ is arbitrarily small. In Section \ref{sec_liq_single}, we study the optimal liquidation for the one-asset case and due to the simplicity of the formulae, in the small size limit, we derive the optimal liquidation time explicitly. Using the same equation for the available volume we study a linearly liquidated portfolio of assets in Section \ref{sec_portfolio_liq}. Finally, Section \ref{sec_numerical_results} is devoted to the numerical results where we applied our algorithm to a test portfolio of corporate bonds, the parameters of which are computed using real market data.

\section{Framework}\label{sec_framework}

In this section, we first recall the framework of the Locally Linear Order Book (LLOB for short) model introduced in \cite{toth2011anomalous}. We then show how to use it to assess the costs of liquidation, in particular on OTC markets, for example, the corporate bonds market. Even though there is no order book for the corporate bond market, electronic trading platforms form a rough approximation of it. It is as if there exists an unobservable order book, hereafter called a latent order book, and that one can observe block prices as a function of the price-volume dynamics of this order book. 
\subsection{The Locally Linear Order Book model}\label{sec_llob}

Initially the LLOB model emerged from an empirical fact that in limit order books the latent volumes around the best ticks are linear in price deviation from the best price, even if it is not directly reflected in the order book. In this section we describe the LLOB model in the initial context of order driven markets. The general idea of the LLOB is that there exists a latent order book which, at any time~$t$, aggregates the total intended volume to be potentially sold at price $p>0$ or above $\mathcal{V}_+(t,p)$ and the total intended volume to be potentially bought at price $p$ or below $\mathcal{V}_-(t,p)$. The latent volumes $\mathcal{V}_+(t,p)$ and $\mathcal{V}_-(t,p)$ are not the volumes revealed in the observable order book but the volumes that would be revealed as limit or market orders if the price comes closer to $p$ at some point (in short, as stated in \cite{toth2011anomalous}, the latent volumes reflect intentions that do not necessarily materialize). \\

Between $t$ and $t+dt$, new buy and sell orders of unit volume may arrive at levels $p_t \mp u$ where $u>0$, with corresponding intensity rates $\lambda(u)$. At the same time the buyers and sellers who have already sent orders at $p_t\mp u$ might want to change the price to $p_t \mp u'$, for $u'>0$, at rate $\nu(u,u')$, or even cancel an order temporarily in the case $u'=+\infty$. \\

Let us assume that the price process $p_t$ is a Brownian motion, which may not be well suited to order books due to microstructural effects, but is suitable to approximate the price process on OTC markets. We assume that either $u'=+\infty$ with rate $\nu_\infty (u)$ or that the change of price is a Brownian motion. We define $D(u)=\int_0^{+\infty} (u-u')^2 \nu(u,u') du'$ interpretable as the squared volatility of intentions. Let us denote by $\rho_{\pm}(t,u)$ a latent volume averaged over price paths, the equation for which is
\begin{align*}
\begin{cases}
& \partial_t \rho_{\pm}(t,u) = \frac{1}{2}\partial_{uu}^2 \big(\mathcal{D}(u)\rho_{\pm}(t,u)\big)-\nu_\infty (u)\rho_{\pm}(t,u) + \lambda(u), \\
& \rho_{\pm}(t,u) = 0 \text{ for } (t,u)\in \mathbb{R}_+ \times \mathbb{R}_-,
 \end{cases}
\end{align*}
where $\mathcal{D}(u)=D(u)+\sigma^2$ and $\sigma>0$ is the daily price volatility.\\

For arbitrary functions $D(u),\lambda(u),\nu_\infty(u)$ the explicit form of the stationary solution of the above PDE is not known. However, in the case where new orders appear uniformly, i.e $\lambda(u)=\lambda$ and $\mathcal{D}(u)=\mathcal{D}$ independent of $u$, the exact stationary solution is
\begin{align}\tag{$\rho$}\label{eq_exact_sol_LLOB}
 \rho(u)= \rho_{\infty}\Big( 1 - \exp\big(-\frac{u}{u^\star}\big) \Big),
\end{align}
where $\rho_{\infty}=\frac{\lambda}{\nu_\infty},u^\star = \sqrt{\frac{\mathcal{D}}{2\nu_\infty}}$. The function $\rho(u)$ is the density of order book trades, or more precisely for all $u'>0$ $\rho(u')=\frac{dV}{du}(u')$, where $V$ is the order book volume as a function of the distance from the mid-price. The meaning of $u^\star$ is the width of the linear price change zone: for $u\ll u^\star$, that is for small deviations from the mid-price, the price depends linearly on the volume, whereas it stays constant for $u\gg u^\star$. The constant $\rho_\infty$ is understood to be the density of an order book trade far away from the mid-price $p_t$. Precisely, it is the inverse of the asymptotic large size market elasticity.\footnote{The asymptotic large size market elasticity is the incremental price needed to trade an incremental volume when trading volume is large relative to ADV.} We define the asymptotic market elasticity as $\epsilon_{\text{asympt.}} = \frac{1}{\rho_\infty}$, and the ``naive'' market elasticity as $\epsilon_{\text{naive}} = \frac{\sigma}{\text{ADV}}$.
 \\

Let us consider a buy order. We integrate Equation \eqref{eq_exact_sol_LLOB} over $u$ from mid-price to~$\Delta p$. The resulting equation gives the order book volume as a function of price change:
\begin{align*}
 V(\Delta p)= \rho_\infty \Big(\Delta p -u^\star \big(1-e^{-\frac{\Delta p}{u^\star}}\big)\Big).
\end{align*}
In the case $\Delta p \ll u^\star$, the price impact varies as the square root of the trade size. For large trade sizes, the price impact is a linear function of the trade size, imitating the increasing cost of trading when the traded volume is bigger than one the market can digest. \\

Note that the above model corresponds to trades that can be done in a single day, and is considered as a one day model. We consider a ``linear'' liquidation in which the block of assets is unwound in equal parts over a number of days $T$ which needs to be determined. If we denote the total block size as~$N$ and a daily trade size of~$\frac{N}{T}$, the cost of trading each block is given by
\begin{align*}
 C_{\text{block}}(T)=\frac{N}{T} \Delta p \Big(\frac{N}{T}\Big),
\end{align*}
where $\Delta p(v)$ is defined as the solution of $V(w)=v$ for $w\in \mathbb{R}, v\in \mathbb{R}_+$. The total direct costs are given by the sum of $C_{\text{block}}$ over the number of days, which is
\begin{align*}
 \text{DC}(T) = C_{\text{block}}(T) T = N\Delta p \Big(\frac{N}{T}\Big).
\end{align*}
We introduce the following parametrization for the quantity $\rho_\infty$:
\begin{align*}
 \rho_\infty = \alpha_\infty \frac{\text{ADV}}{\sigma},
\end{align*}
to be compared with $\rho_\infty = \frac{\lambda}{\nu_\infty}$ in \cite{toth2011anomalous}. In other words, we take the number of daily orders $\lambda$ (of unit volume) equal to the average daily number of unit volumes (equal to ADV), and the rate at which buy or sell orders are canceled equal to~$\frac{\sigma}{\alpha_\infty}$ where $\alpha_\infty>0$ is a free parameter. We can rewrite the dimensionless parameter $\alpha_\infty$ in terms of market elasticity so that
\begin{align*}
    \alpha_\infty = \frac{\frac{\sigma}{\text{ADV}}}{\frac{1}{\rho_\infty}}=\frac{\epsilon_{\text{naive}}}{\epsilon_{\text{asympt.}}},
\end{align*}
So the physical meaning of and intuition behind this is the ratio of the ``naive'' market elasticity for trades not large compared with ADV to the value of market elasticity for trades materially larger than ADV. We also assume that the width of the linear region is of the order of one day's price move, so that $u^\star = \sigma$. 

\subsection{Single asset liquidation}\label{sec_liq_single}

To account for a trading firm's risk aversion, we consider a running penalty proportional to the standard deviation of the PnL of the entire block liquidation, which is a measure of risk usually used in practice. As we assume a linear liquidation schedule, the penalty can be written
\begin{align*}
 \phi(T)= \gamma \sqrt{\mathbb{V}(PnL)}= \frac{\gamma}{\sqrt{3}}P_0 N \sigma \sqrt{T}
\end{align*}
where $\gamma$ is a number of standard deviations of the PnL representing the risk tolerance of the firm and $P_0$ is taken to be the bond price (in units in which par is $1$) and $N$ is the face amount. The effect of the volatility penalty is to incentivize the optimizer to not take too much time with liquidating the position. \\

Let us consider the case $\Delta p \ll u^\star$. By taking a Taylor expansion of Equation \eqref{eq_exact_sol_LLOB} at $\frac{\Delta p}{u^\star}=0$, we have
\begin{align*}
 \Delta p (V) = \sqrt{\frac{2V u^\star}{\rho_\infty}}. 
\end{align*}
This is the commonly assumed square root law for price penalty as a function of volume, see \cite{bouchaud2010price}, for example. Given an assumed linear liquidation schedule the cost for each block is therefore given by
\begin{align*}
 C_{\text{block}}= \frac{N}{T}\Delta p_{\text{block}}=\sqrt{\frac{2 u^\star}{\rho_\infty}} \Big(\frac{N}{T}\Big)^{3/2}.
\end{align*}
The total cost is a sum of costs of all blocks and the volatility penalty
\begin{align*}
 \text{TC}(T) = \text{DC}(T) + \phi(T) = \sqrt{\frac{2 u^\star}{\rho_\infty}} \frac{N^{3/2}}{\sqrt{T}}+ \frac{\gamma P_0 N \sigma \sqrt{T}}{\sqrt{3}}.
\end{align*}
Let us express this money amount in terms of a cost per bond:
\begin{align*}
 c(T)= \frac{\text{TC(T)}}{N} = \sqrt{\frac{2 N u^\star}{T\rho_\infty}} + \frac{\gamma P_0 \sigma \sqrt{T}}{\sqrt{3}}. 
\end{align*}
We now solve the optimal liquidation problem by setting the first derivative of $c$ with respect to $T$ equal to zero. Computations lead to the following optimal liquidation time:
\begin{align*}
 T^\star = \frac{\sqrt{3 N}}{\gamma P_0 \sigma}\sqrt{\frac{2u^\star}{\rho_\infty}},
\end{align*}
which provides the cost per bond:
\begin{align*}
    c(T^\star) = 2\Big(\frac{2u^\star}{\rho_\infty}\Big)^{1/4}\frac{\sqrt{\gamma P_0\sigma}N^{1/4}}{3^{1/4}}.
\end{align*}
\begin{remark}
By setting $u^\star=\sigma,\alpha_\infty=1$  we obtain 
\begin{align*}
 \tilde{T}^\star = \frac{\sqrt{6}}{\gamma P_0}\sqrt{\frac{N}{\text{ADV}}}, \quad \tilde{c}(T^\star) = \frac{2^{5/4}\sqrt{\gamma P_0}\sigma}{3^{1/4}}\Big(\frac{N}{\text{ADV}}\Big)^{1/4}.
\end{align*}

A similar analysis to the above may be done with the large trade size limit:
\begin{align*}
 \lim_{N\rightarrow +\infty} c^\star = 3^{-1/3}(2^{-2/3}+2^{1/3})(\gamma P_0)^{1/3}\Big(\frac{N}{\text{ADV}}\Big)^{1/3},
\end{align*}
So in between the small and large limits the cost dependence on trade size changes from $N^{1/4}$ to $N^{1/3}$. 
\end{remark}

In the following section, we show how to extend this framework to the multi-asset case. 

\subsection{Portfolio liquidation}\label{sec_portfolio_liq}

When moving to the multi-asset version of the optimization, one needs to create multi-asset versions of both the direct cost and the volatility penalty. The problem of cross-impact emerging when, for example, trades of a certain amount of one asset influence the price of another asset, is not treated in our model. Optimal liquidation models taking into account cross-impact (see, for example,\cite{mastromatteo2017trading}) exist, however it is hard to estimate cross-impact matrices in OTC markets, notably due to fragmentation. For the sake of simplicity, we assume that the total direct cost is the sum of the individual direct costs. \\

The multi-asset volatility penalty for the portfolio is a straightforward extension of the single asset version. It is an integral over time of the covariances of the remaining positions. The position function $N^i(t)$ for the bond $i$ that is linearly liquidated over time is given by
\begin{align*}
 N^i(t)=N_0^i\left(1-\frac{t}{T^{\star i}}\right)_+,
\end{align*}
where $N_0^i\in \mathbb{R}$ is the initial position in the bond $i$, and $T^{\star i}$ refers to the final liquidation time, such that $N^i(T^{\star i}) = 0$. The total variance of the PnL can be expressed as a sum over covariance terms
\begin{align*}
 \sum_{i,j=1}^d\!\! \sigma^i\sigma^j \rho^{i,j}N_0^iN_0^j\! \int_{0}^{\min(T^{\star i},T^{\star j})}\!\!\!\left(1\!-\!\frac{t}{T^{\star i}}\right)\!\!\left(1\!-\!\frac{t}{T^{\star j}}\right) dt\!\! = \!\!
 \sum_{i,j=1}^d\!\! \frac{\sigma^i\sigma^j \rho^{i,j}N_0^iN_0^j}{2} \min(T^{\star i},T^{\star j}) \Big(1\!-\!\frac{\min(T^{\star i}\!\!\!\!,T^{\star j})}{3\max(T^{\star i},T^{\star j})}\! \Big). 
\end{align*}

\begin{remark}[Calibration of $\alpha_\infty$]
A simple and intuitive approach for fixing the value of $\alpha_\infty$ can be found by looking at the small size asymptotic limit formula for the optimal liquidation time:
\begin{align*}
 T^{\star}_{asympt}= \frac{\sqrt{6}}{\gamma P_0 \sqrt{\alpha_\infty}}\sqrt{\frac{N}{\text{ADV}}},
\end{align*}
where $\alpha_\infty$ has been reintroduced. Let us assume a bond priced at par, and impose the condition $T_{asympt}^\star(N=\text{ADV})=1$, implying that it is reasonable to trade the ADV in one day, therefore
\begin{align*}
 \alpha_\infty = \frac{6}{\gamma^2}.
\end{align*}
\end{remark}

\section{Numerical results}\label{sec_numerical_results}

In this section we present numerical examples of optimal liquidation using our methodology. We first show an application to a long-short portfolio of two correlated bonds sharing same characteristics except that one is much more liquid than the other. Then, we present the results obtained on a long-short portfolio of $20$ bonds. In all the numerical results, we choose a risk aversion parameter $\gamma = 0.5$.

\subsection{Long-short portfolio with two correlated bonds}

This test case demonstrates the disadvantages of a line by line liquidation of a long/short portfolio, typically used in vendor liquidity stress testing offerings. The test portfolio consists of two bond positions of the same size ($27$ and $-27$ respectively), where the bonds have same price of $141.49\$$ and $7\%$ annualized volatility, but different ADVs: 30 for the first bond and 3 for the second, so that the first one is more liquid.\\

A liquidation strategy based on individual liquidation would result in the more liquid bond being unwound rapidly and the less liquid one slower. But clearly the optimal way to liquidate this portfolio is to unwind these positions with the same liquidation strategy, especially the same timescale. This would minimize total PnL variance thereby allowing for a longer liquidation time and less costs.
\vspace{-5mm}
\begin{figure}[H]
\begin{minipage}{.48\textwidth}
\centering
  \includegraphics[width=0.99\textwidth]{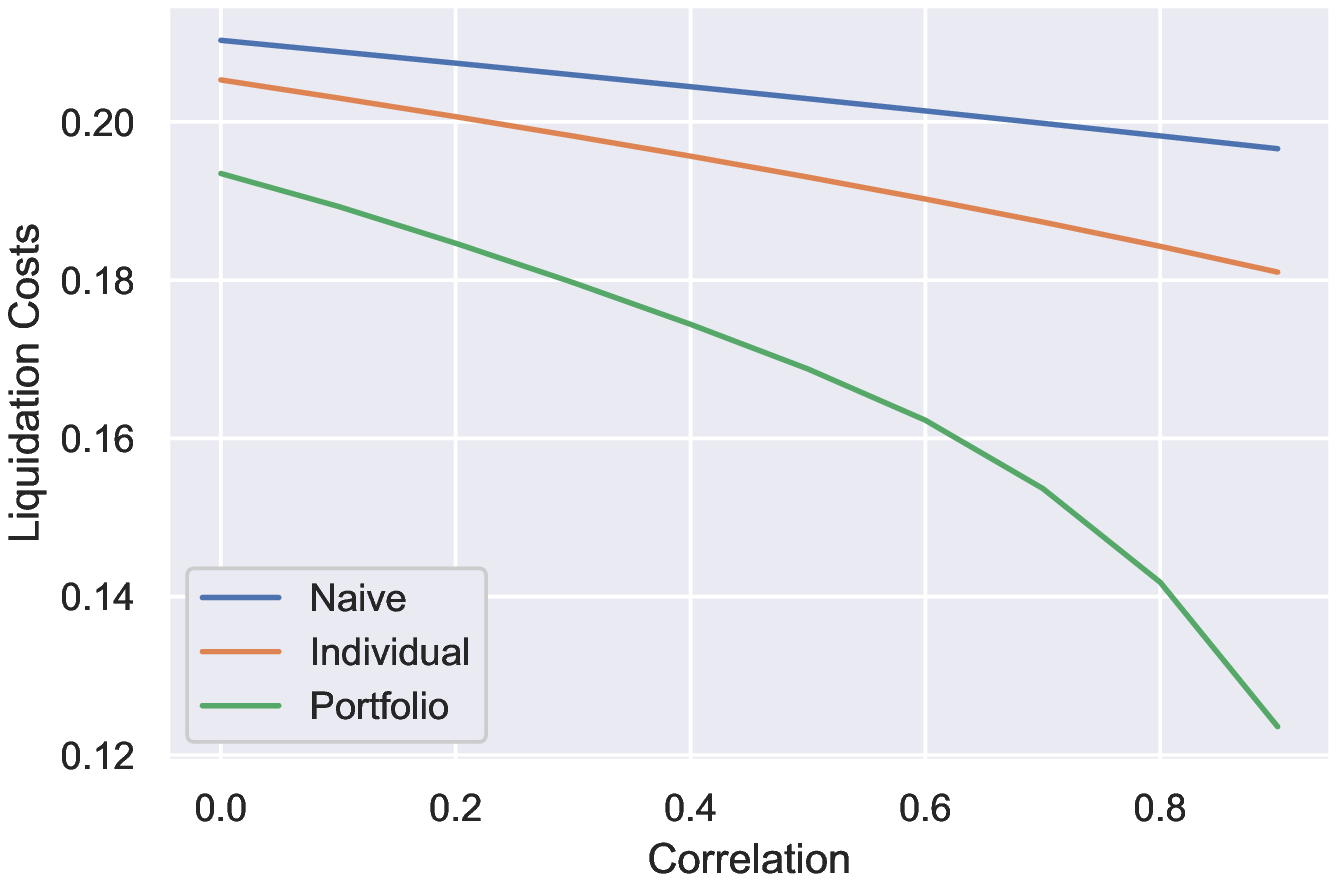}
  \vspace{-3mm}
  \caption{Optimal portfolio liquidation costs with respect to correlation for different strategies.}\label{costs_vs_corr}
\end{minipage}%
\quad
\begin{minipage}{.48\textwidth}
\centering
    \includegraphics[width=0.99\textwidth]{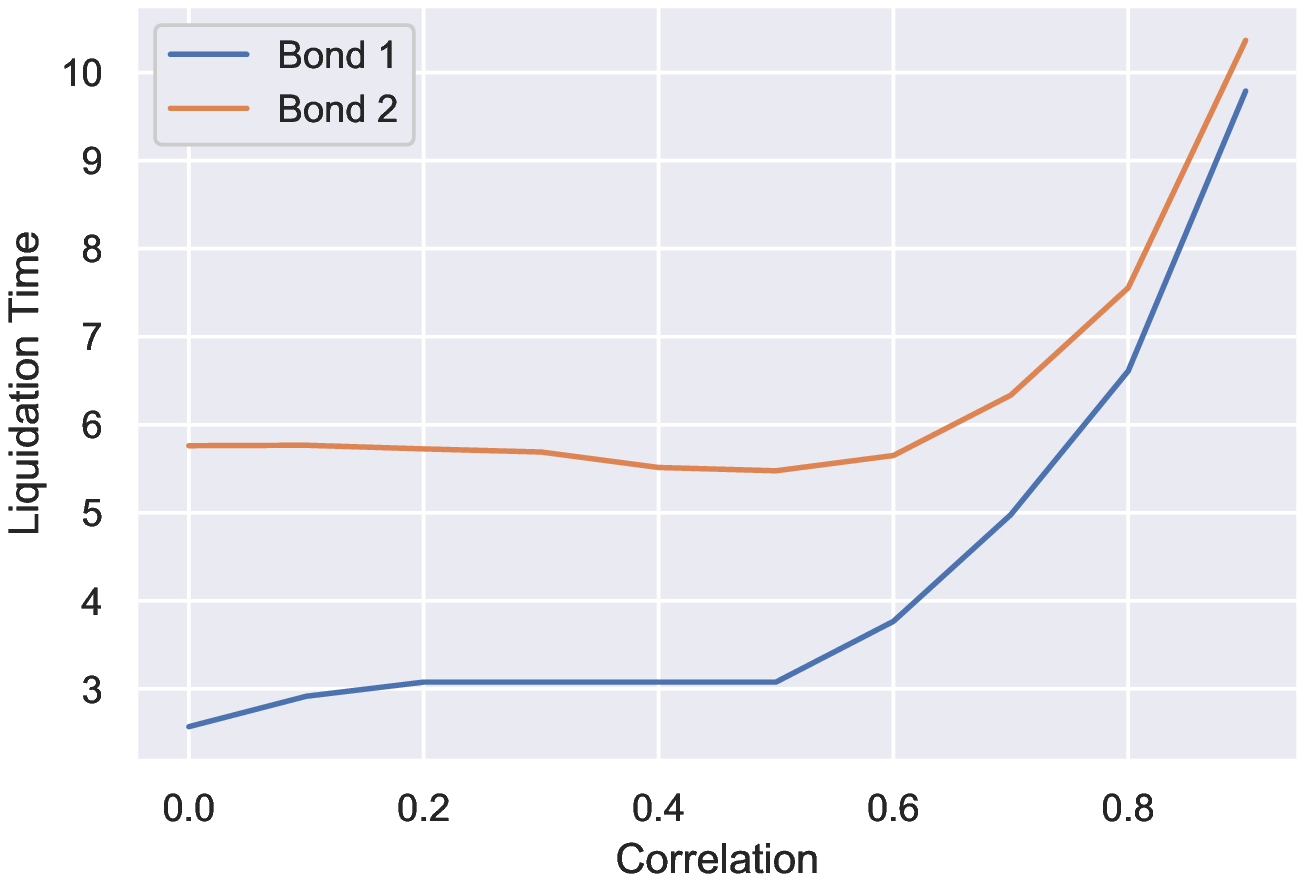}
  \vspace{-3mm}
  \caption{Liquidation times for 2-bond case.}\label{lq_times_corr}
  \vspace{6mm}
\end{minipage}
\end{figure}
In Figure \ref{costs_vs_corr}, we show the liquidation costs as a function of the correlation between the two bond price returns. We refer to individual optimization to be the line by line liquidation of the positions, each bond is liquidated independently of the others. Such individual liquidation implies that the liquidation cost for the portfolio is the sum of the liquidation for each bond, including the standard deviation penalty. By a naive strategy we refer to a strategy with $T_{naive}^{\star i}=\frac{N^i}{\text{ADV}^i}$. \\

The liquidation costs are decreasing functions of the correlation for all strategies. The difference between portfolio optimization costs and individual optimization costs even for small correlations is the consequence of the choice of the penalty function, and we have no intention to compare the values directly. We are mostly interested to compare the costs dependence on the correlation level. Notably, in Figure \ref{costs_vs_corr}, we see that the costs of portfolio liquidation are decreasing more steeply when the correlation level increases compared to the individual optimization. Before the $60\%$ correlation level the decrease in costs is mostly linear for portfolio optimization, and for the correlation levels above $60\%$ it becomes more concave. In this specific case, the line by line liquidation provides lower costs than the naive liquidation. However in general, this has no reason to be true.\\

In Figure \ref{lq_times_corr}, we show the liquidation times for 2 bonds as a function of correlation in the case of portfolio optimization. Below $45\%$ correlation, the optimal liquidation times appear to be almost independent of correlation (3 days for the first bond and 6 days for the second bond). Then the liquidation time of the less liquid bond decreases so as to approach the liquidation time of the liquid bond. As correlation increases, both times increase, converging to the same optimal time of 10 days.  


\subsection{Long-short portfolio of 20 bonds }

We have chosen a set of $20$ random bonds from the USD Investment Grade and High Yield universes, with somewhat random position sizes assigned. In Table \ref{fig_description_bonds} we show the main characteristics of the portfolio and in Table \ref{corrmatr} its correlation matrix.\\

We can summarize the bonds' parameters as (from 3rd April 2020 unless otherwise noted):
\begin{itemize}
    \item The gross value is about $\$40$M: $\$25$M long and $\$15$M short.
    \item ADV is estimated available daily volume calibrated on TRACE volume data and varies from $\$2$M to $\$21$M per day across the 20 bond in this portfolio.
    \item Volatility is 22 business days (one month) historical volatility and varies from $3.8\%$ to $63\%$ annual.
    \item Bid-ask is set to 20bp to provide a minimum level for all bonds.
\end{itemize}

For the particular portfolio chosen and for the time period chosen (three months to early June 2020) the average correlation between bonds was about 20\%. However, this may not be representative of typical bonds in typical time periods. It is therefore worth looking at the behavior of optimal liquidation cost and time versus correlation.\\

\vspace{-10mm}
\begin{table}[H]
\renewcommand\arraystretch{0.94}
\centering
\begin{tabular}{lllll}
\hline
\multicolumn{1}{c}{Bond} & \multicolumn{1}{c}{\begin{tabular}[c]{@{}c@{}}ADV \\ \$M/day\end{tabular}} & \multicolumn{1}{c}{\begin{tabular}[c]{@{}c@{}}min \\ bid-ask\end{tabular}} & \multicolumn{1}{c}{\begin{tabular}[c]{@{}c@{}}Annual. \\ vol\end{tabular}} & \multicolumn{1}{c}{\begin{tabular}[c]{@{}c@{}}Bond face\\ amount  \$M\end{tabular}}\\ \hline
1 & 3.0 & 0.20 \% & 7.0 \% & 27 \\
2 & 3.0 & 0.20 \% & 8.8 \% & -33 \\ 
3 & 8.0 & 0.20 \% & 12.5 \% & -24 \\
4 & 2.5 & 0.20 \% & 4.9 \% & -31.5 \\ 
5 & 3.5 & 0.20 \% & 13.0 \% & 27 \\ 
6 & 6.0 & 0.20 \% & 7.1 \% & -2 \\
7 & 4.5 & 0.20 \% & 21.5 \% & -1.5 \\
8 & 2.0 & 0.20 \% & 18.4 \% & -1 \\ 
9 & 5.0 & 0.20 \% & 3.8 \% & -1 \\ 
10 & 2.5 & 0.20 \% & 11.1 \% & -0.71 \\
11 & 5.0 & 0.20 \% & 32.9 \% & 42 \\
12 & 3.0 & 0.20 \% & 13.0 \% & -42 \\ 
13 & 4.5 & 0.20 \% & 11.3 \% & 40 \\ 
14 & 17.5 & 0.20 \% & 11.8 \% & -40 \\ 
15 & 21.5 & 0.20 \% & 10.8 \% & 37.5 \\ 
16 & 20.5 & 0.20 \% & 63.4 \% & 2 \\ 
17 & 2.5 & 0.20 \% & 60.7 \% & 1.5 \\ 
18 & 9.5 & 0.20 \% & 11.7 \% & 1 \\ 
19 & 3.0 & 0.20 \% & 26.4 \% & -1 \\ 
20 & 2.0 & 0.20 \% & 12.9 \% & -0.77 \\ \hline
\end{tabular}
\vspace{-2mm}
\caption{Test portfolio characteristics.}
\label{fig_description_bonds}
\end{table}

\begin{figure}[H]
\centering
\begin{minipage}{.48\textwidth}
  \centering
  \includegraphics[width=0.99\textwidth]{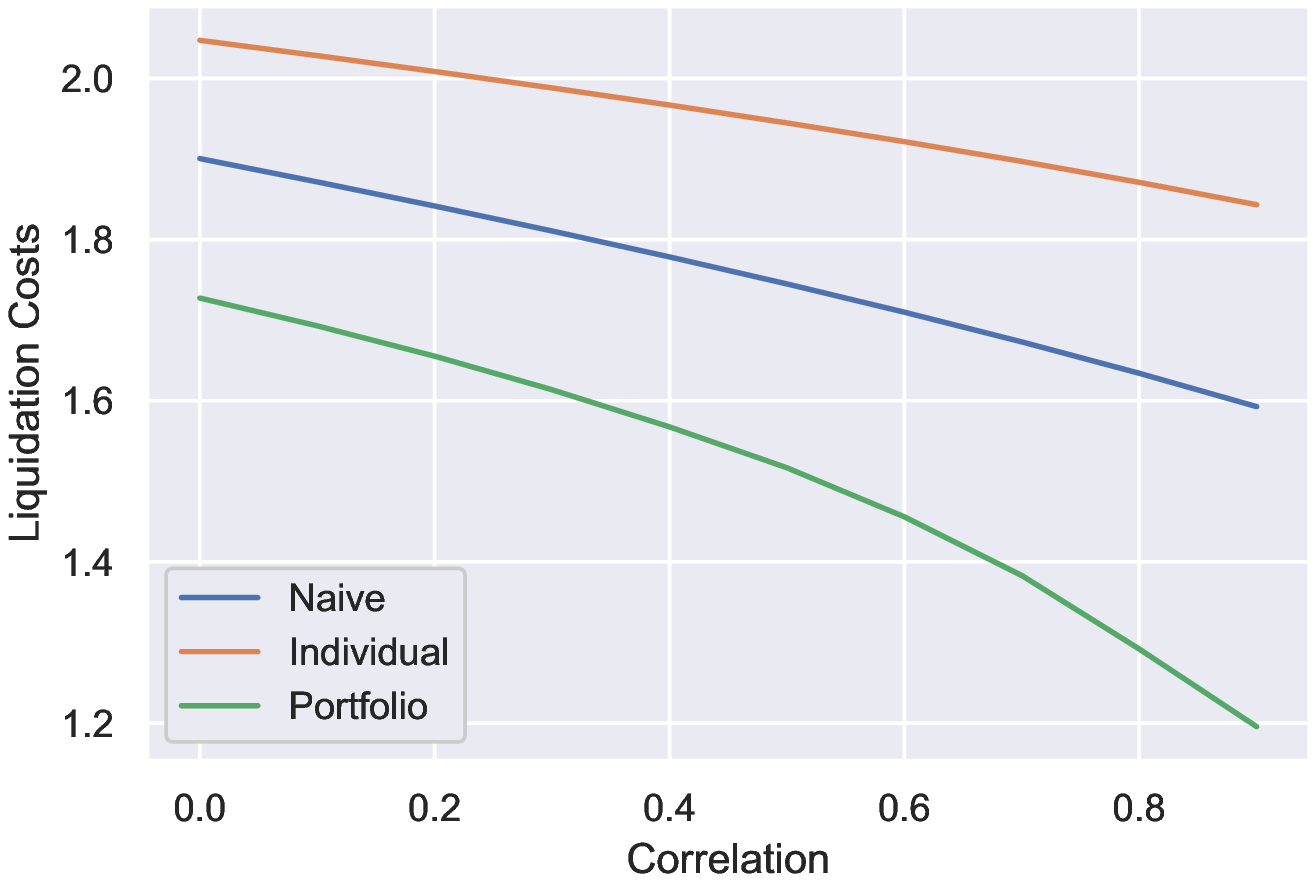}
  \vspace{-3mm}
  \caption{Optimal liquidation costs with portfolio optimizer.}\label{opt_lq_cost}
\end{minipage}%
\quad
\begin{minipage}{.48\textwidth}
  \centering
    \includegraphics[width=0.99\textwidth]{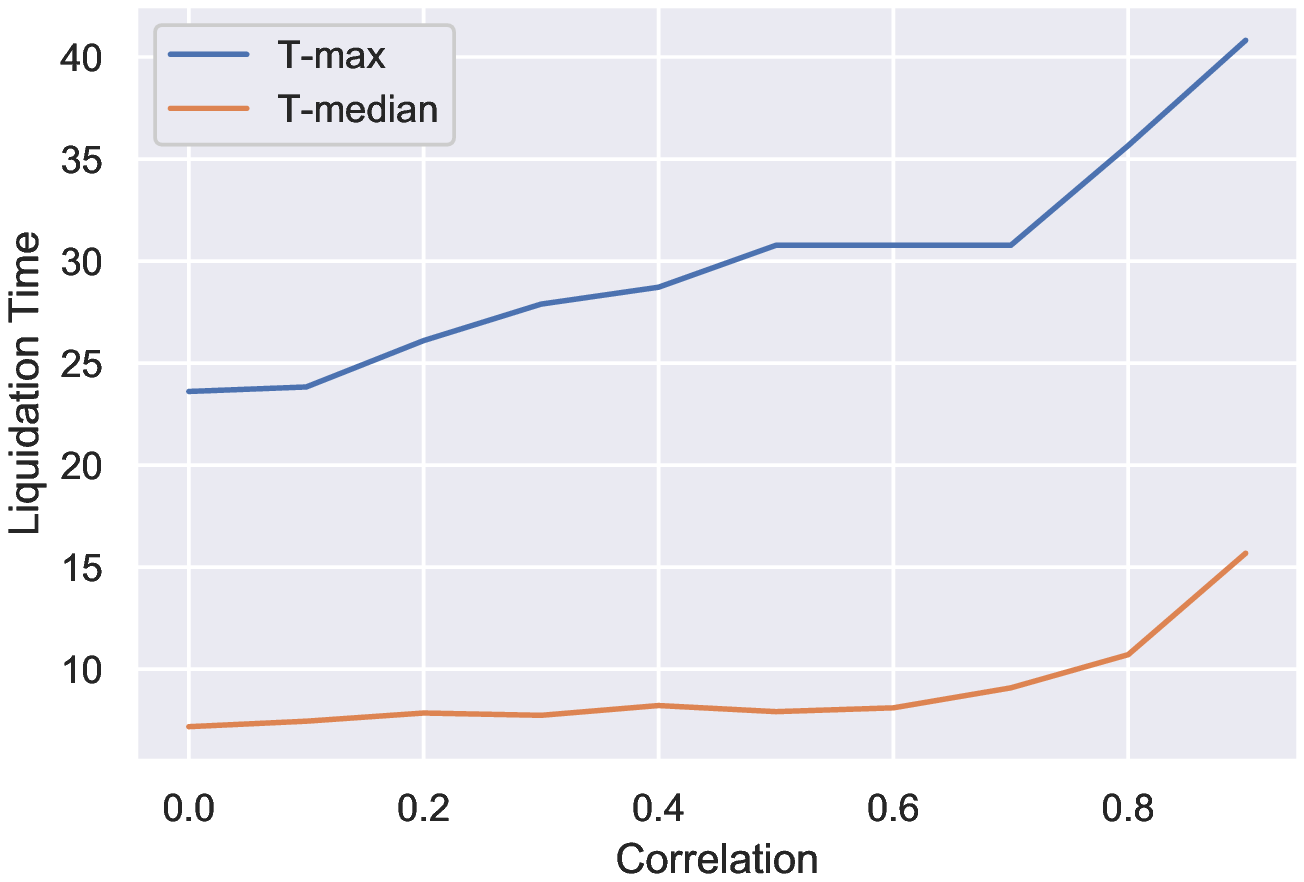}
  \vspace{-3mm}
  \caption{Optimal liquidation times with portfolio optimizer.}\label{opt_lq_time}
\end{minipage}
\end{figure}

In Figure \ref{opt_lq_cost}, we show the optimal portfolio liquidation cost - both the direct cost and the full cost including volatility penalty – versus pairwise correlation, with all off-diagonal correlations set to the same value.
\vspace{-5mm}

As expected, this long-short portfolio has a decreasing liquidation cost as correlation increases. As none of the other models are sensitive to correlation, their direct cost values are constant, though their volatility penalty naturally decreases with increasing correlation when evaluated using the portfolio cost function. As in the 2-bond example we can notice that the decrease of the costs for the portfolio optimization is steeper for all correlation levels compared to other strategies considered. We can also notice that the costs of portfolio optimization becomes concave for the correlation levels above $50\%$. Note that, in this example, the line by line optimization provides higher liquidation costs compared to a simple naive liquidation strategy.  \\

It is also interesting to look at liquidation times for the optimal portfolio liquidation as correlation increases, shown in Figure \ref{opt_lq_time}. The optimizer is taking advantage of the higher correlation causing a reduced volatility penalty for slower liquidation. Both the median and maximum time are increasing monotonously with respect to the correlation level. \\

In all of these optimizations, there have been no constraints, apart from a very high time constraint of 100 days which was never effective. However, this model can be used to calculate an optimal liquidation strategy under time constraint, which is very useful for liquidity stress testing. Taking our example case portfolio, we can see that with a maximum liquidation deadline of 100 days, the median and maximum liquidation times are about 9 and 32 days, with a direct cost of \$0.805m. It is interesting to see the impact on the cost function as we decrease the deadline. \\

In Table~\ref{fig_comparison_deadline}, we compare the short deadline results where a time upper bound was used to constrain the optimizer. In Figure \ref{short_deadline_premium}, we show the excess cost above the optimal liquidation cost due to deadline full liquidation shortening. Even though the median time to liquidate the portfolio underlyings was about $9$ days in the optimal case, shortening the liquidation time cutoff to a maximum of $10$ days only causes a minor increase in cost, but as the deadline becomes shorter the costs increase drastically.

\begin{table}[H]
\begin{minipage}[H]{0.51\linewidth}
\setlength\tabcolsep{3pt}
\centering
\begin{tabular}{llllll}
    \hline
    Deadline & \begin{tabular}[c]{@{}l@{}}Portfolio \\liq.cost\end{tabular} & \begin{tabular}[c]{@{}l@{}}Portfolio \\direct cost\end{tabular} & \begin{tabular}[c]{@{}l@{}}Portfolio \\T-median\end{tabular} & \begin{tabular}[c]{@{}l@{}}Portfolio \\T-max\end{tabular}  \\ \hline
    100 & 1.605 & 0.805  & 9.4 & 32.5\\ 
    20 & 1.607 & 0.812  & 9.4 & 20   \\ 
    15 & 1.612 & 0.829  & 8.9 & 15   \\ 
    10 & 1.635 & 0.903  & 8.2 & 10   \\ 
    7.5 & 1.677 & 1.004  & 7.5 & 7.5  \\ 
    5 & 1.780 & 1.176  & 5 & 5   \\ 
    3 & 1.999 & 1.503  & 3 & 3  \\ 
    2 & 2.282 & 1.875  & 2 & 2 \\ 
    1 & 3.061 & 2.772  & 1 & 1  \\ \hline
    \end{tabular}
    \caption{Short deadline costs and times comparison.}
    \label{fig_comparison_deadline}
\end{minipage}\hfill
\begin{minipage}{0.49\linewidth}
\quad
\includegraphics[width=\textwidth]{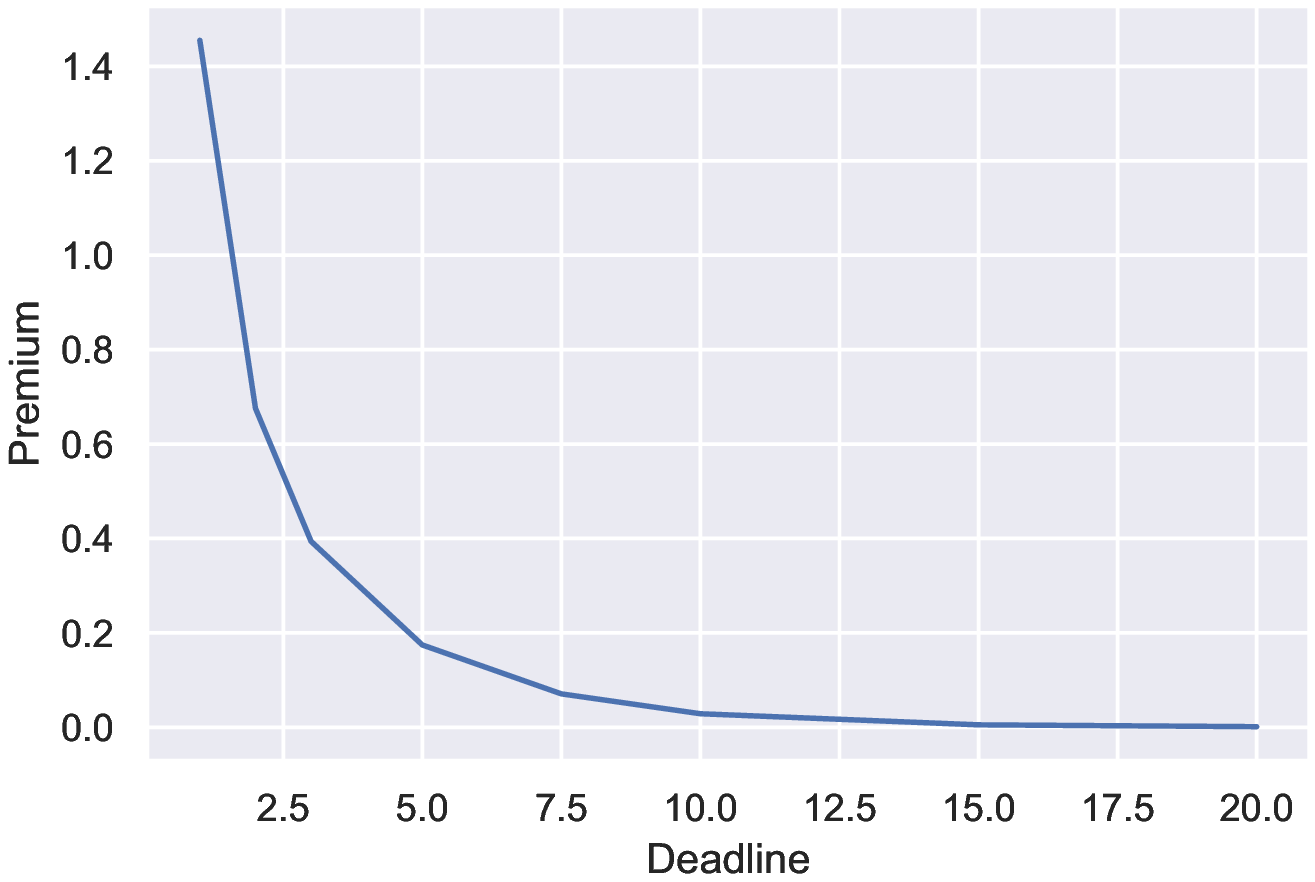}
\vspace{-8mm}
\captionof{figure}{Short deadline premium.}
\label{short_deadline_premium}
\end{minipage}
\end{table}

In Table \ref{fig_comparison_liquidation}, we compare liquidation costs for naive, individual and portfolio optimization strategies and present the median and the maximum liquidation times for the portfolio optimization across different portfolios. The first portfolio corresponds to the test portfolio considered above with the correlation matrix presented in Table \ref{corrmatr}, and other portfolios are the ones with all correlations set to a certain level.\\

For every level of correlation and for the example of $20$ bonds described in Table \ref{fig_description_bonds}, we present the liquidation costs in the naive, individual and portfolio optimization case. For the last case, we also present the direct costs, the median and maximum liquidation time.  
\begin{table}[H]
\begin{center}
\begin{tabular}{lllllllll}
\hline
\multicolumn{1}{l}{Correlation} & \begin{tabular}[c]{@{}l@{}}Naive\\liq. cost\end{tabular} & \begin{tabular}[c]{@{}l@{}}Individual \\liq. cost\end{tabular}  &  \begin{tabular}[c]{@{}l@{}}Portfolio \\liq. cost\end{tabular}  & \begin{tabular}[c]{@{}l@{}}Portfolio \\direct cost\end{tabular}  & \begin{tabular}[c]{@{}l@{}}Portfolio \\T-median\end{tabular} & \begin{tabular}[c]{@{}l@{}}Portfolio \\T-max\end{tabular} \\ 
\hline
\multicolumn{1}{l}{Test} & 1.82 & 1.99 & 1.61   & 0.81   &  9.43 & 32.51  \\ 
\multicolumn{1}{l}{0\%} & 1.90& 2.05 & 1.73   & 0.86   &  7.19 & 23.45  \\ 
\multicolumn{1}{l}{10\%}& 1.87& 2.03 & 1.69   & 0.84   & 7.66 & 24.97  \\ 
\multicolumn{1}{l}{20\%}& 1.84& 2.01 & 1.65   & 0.82   &  8.22 & 25.67  \\ 
\multicolumn{1}{l}{30\%}& 1.81& 1.99 & 1.61   & 0.80   &  8.06 & 27.88    \\ 
\multicolumn{1}{l}{40\%}& 1.78& 1.97 & 1.57   & 0.78   &  8.36 & 29.22   \\ 
\multicolumn{1}{l}{50\%}& 1.74& 1.94 & 1.52   & 0.76   &  7.99 & 27.37  \\ 
\multicolumn{1}{l}{60\%}& 1.71& 1.92 & 1.46   & 0.74   &  8.73 & 29.86  \\ 
\multicolumn{1}{l}{70\%}& 1.67& 1.90 & 1.39   & 0.70   &  9.28 & 30.77  \\ 
\multicolumn{1}{l}{80\%}& 1.63& 1.87 & 1.30   & 0.67   &  10.45 & 36.20  \\ 
\multicolumn{1}{l}{90\%}& 1.59& 1.84 & 1.20   & 0.64   & 11.76  & 40.62    \\ \hline
\end{tabular}
 \caption{Comparison between three type of liquidation.}
 \label{fig_comparison_liquidation}
\end{center}
\end{table}
\vspace{-3mm}

The methodology presented in this paper allows one to obtain the optimal liquidation strategy for a portfolio of bonds in time proportional to $O(d^2)$ where $d$ is the number of bonds. For the test portfolio example, the method works in 5 seconds and for a portfolio of $1000$ bonds, it takes less than $6$ hours, which is reasonable in the context of liquidity stress testing. 

\section{Conclusion}

In this paper, we presented an optimal portfolio liquidation model based on the Locally Linear Order Book framework with an application to liquidity stress testing on OTC markets. The model has only one free parameter to be calibrated. When the traded volume is small, the optimal liquidation time in the single asset case is obtained analytically and is proportional to the square root of the ratio between the volume being liquidated and the average daily volume. In the case of portfolio liquidation, our simple and reasonably fast optimization procedure established in this paper can be applied.  

\begin{landscape}
\begin{table}[]
\centering
\setlength\tabcolsep{3pt}
\begin{tabular}{
>{\columncolor[HTML]{EFEFEF}}l 
>{\columncolor[HTML]{FFFFFF}}l 
>{\columncolor[HTML]{FFFFFF}}l 
>{\columncolor[HTML]{FFFFFF}}l 
>{\columncolor[HTML]{FFFFFF}}l 
>{\columncolor[HTML]{FFFFFF}}l 
>{\columncolor[HTML]{FFFFFF}}l 
>{\columncolor[HTML]{FFFFFF}}l 
>{\columncolor[HTML]{FFFFFF}}l 
>{\columncolor[HTML]{FFFFFF}}l 
>{\columncolor[HTML]{FFFFFF}}l 
>{\columncolor[HTML]{FFFFFF}}l 
>{\columncolor[HTML]{FFFFFF}}l 
>{\columncolor[HTML]{FFFFFF}}l
>{\columncolor[HTML]{FFFFFF}}l 
>{\columncolor[HTML]{FFFFFF}}l 
>{\columncolor[HTML]{FFFFFF}}l 
>{\columncolor[HTML]{FFFFFF}}l 
>{\columncolor[HTML]{FFFFFF}}l 
>{\columncolor[HTML]{FFFFFF}}l
>{\columncolor[HTML]{FFFFFF}}l }
\cline{2-21}
\cellcolor[HTML]{FFFFFF} & \cellcolor[HTML]{EFEFEF}1 & \cellcolor[HTML]{EFEFEF}2 & \cellcolor[HTML]{EFEFEF}3 & \cellcolor[HTML]{EFEFEF}4 & \cellcolor[HTML]{EFEFEF}5 & \cellcolor[HTML]{EFEFEF}6 & \cellcolor[HTML]{EFEFEF}7 & \cellcolor[HTML]{EFEFEF}8 & \cellcolor[HTML]{EFEFEF}9 & \cellcolor[HTML]{EFEFEF}10 & \cellcolor[HTML]{EFEFEF}11 & \cellcolor[HTML]{EFEFEF}12 & \cellcolor[HTML]{EFEFEF}13 & \cellcolor[HTML]{EFEFEF}14 & \cellcolor[HTML]{EFEFEF}15 & \cellcolor[HTML]{EFEFEF}16 & \cellcolor[HTML]{EFEFEF}17 & \cellcolor[HTML]{EFEFEF}18 & \cellcolor[HTML]{EFEFEF}19 & \cellcolor[HTML]{EFEFEF}20 \\ \cline{2-21}
\multicolumn{1}{l|}{\cellcolor[HTML]{EFEFEF}1} & 1.00 & 0.10 & 0.12 & 0.14 & 0.14 & 0.16 & 0.42 & 0.30 & 0.30 & 0.10 & 0.16 & -0.11 & 0.32 & 0.35 & 0.48 & 0.24 & -0.01 & 0.52 & 0.24 & 0.43 \\ 
\multicolumn{1}{l|}{\cellcolor[HTML]{EFEFEF}2} & 0.10 & 1.00 & 0.30 & -0.23 & 0.16 & 0.26 & -0.03 & 0.23 & 0.04 & 0.26 & 0.30 & 0.24 & 0.29 & 0.04 & -0.09 & 0.22 & 0.19 & 0.11 & 0.36 & 0.25 \\ 
\multicolumn{1}{l|}{\cellcolor[HTML]{EFEFEF}3} & 0.12 & 0.30 & 1.00 & -0.05 & -0.17 & 0.59 & 0.28 & 0.29 & 0.22 & -0.02 & 0.26 & 0.11 & 0.40 & 0.33 & 0.29 & 0.39 & 0.14 & 0.24 & 0.05 & 0.05 \\ 
\multicolumn{1}{l|}{\cellcolor[HTML]{EFEFEF}4} & 0.14 & -0.23 & -0.05 & 1.00 & 0.23 & 0.06 & 0.14 & 0.11 & 0.37 & -0.13 & 0.05 & 0.31 & 0.22 & 0.30 & 0.31 & 0.25 & 0.36 & 0.25 & 0.25 & 0.19 \\ 
\multicolumn{1}{l|}{\cellcolor[HTML]{EFEFEF}5} & 0.14 & 0.16 & -0.17 & 0.23 & 1.00 & 0.15 & -0.31 & 0.34 & 0.09 & 0.12 & -0.00 & 0.24 & 0.36 & -0.03 & 0.07 & 0.13 & -0.05 & -0.02 & 0.23 & -0.01 \\ 
\multicolumn{1}{l|}{\cellcolor[HTML]{EFEFEF}6} & 0.16 & 0.26 & 0.59 & 0.06 & 0.15 & 1.00 & -0.00 & 0.40 & 0.46 & 0.22 & 0.15 & 0.13 & 0.35 & 0.25 & 0.13 & 0.23 & -0.00 & 0.23 & 0.13 & 0.14 \\ 
\multicolumn{1}{l|}{\cellcolor[HTML]{EFEFEF}7} & 0.42 & -0.03 & 0.28 & 0.14 & -0.31 & -0.00 & 1.00 & 0.08 & 0.39 & 0.13 & 0.03 & 0.01 & 0.17 & 0.08 & 0.08 & 0.09 & 0.09 & 0.27 & 0.03 & 0.28 \\ 
\multicolumn{1}{l|}{\cellcolor[HTML]{EFEFEF}8} & 0.30 & 0.23 & 0.29 & 0.11 & 0.34 & 0.40 & 0.08 & 1.00 & 0.48 & 0.36 & 0.10 & -0.08 & 0.13 & 0.06 & 0.15 & 0.20 & 0.02 & 0.25 & 0.35 & 0.12 \\ 
\multicolumn{1}{l|}{\cellcolor[HTML]{EFEFEF}9} & 0.30 & 0.04 & 0.22 & 0.37 & 0.09 & 0.46 & 0.39 & 0.48 & 1.00 & 0.20 & 0.07 & 0.24 & 0.28 & 0.23 & 0.12 & 0.05 & 0.14 & 0.34 & 0.06 & 0.34 \\ 
\multicolumn{1}{l|}{\cellcolor[HTML]{EFEFEF}10} & 0.10 & 0.26 & -0.02 & -0.13 & 0.12 & 0.22 & 0.13 & 0.36 & 0.20 & 1.00 & -0.05 & 0.19 & 0.05 & -0.04 & -0.12 & -0.08 & 0.05 & 0.13 & 0.14 & 0.32 \\ 
\multicolumn{1}{l|}{\cellcolor[HTML]{EFEFEF}11} & 0.16 & 0.30 & 0.26 & 0.05 & -0.00 & 0.15 & 0.03 & 0.10 & 0.07 & -0.05 & 1.00 & 0.12 & 0.22 & 0.29 & 0.29 & 0.12 & 0.27 & 0.19 & 0.23 & 0.23 \\ 
\multicolumn{1}{l|}{\cellcolor[HTML]{EFEFEF}12} & -0.11 & 0.24 & 0.11 & 0.31 & 0.24 & 0.13 & 0.01 & -0.08 & 0.24 & 0.19 & 0.12 & 1.00 & 0.39 & 0.18 & 0.05 & 0.11 & 0.52 & 0.03 & 0.08 & 0.25 \\ 
\multicolumn{1}{l|}{\cellcolor[HTML]{EFEFEF}13} & 0.32 & 0.29 & 0.40 & 0.22 & 0.36 & 0.35 & 0.17 & 0.13 & 0.28 & 0.05 & 0.22 & 0.39 & 1.00 & 0.40 & 0.40 & 0.49 & 0.12 & 0.39 & 0.13 & 0.38 \\ 
\multicolumn{1}{l|}{\cellcolor[HTML]{EFEFEF}14} & 0.35 & 0.04 & 0.33 & 0.30 & -0.03 & 0.25 & 0.08 & 0.06 & 0.23 & -0.04 & 0.29 & 0.18 & 0.40 & 1.00 & 0.82 & 0.48 & 0.25 & 0.69 & 0.38 & 0.57 \\ 
\multicolumn{1}{l|}{\cellcolor[HTML]{EFEFEF}15} & 0.48 & -0.09 & 0.29 & 0.31 & 0.07 & 0.13 & 0.08 & 0.15 & 0.12 & -0.12 & 0.29 & 0.05 & 0.40 & 0.82 & 1.00 & 0.58 & 0.20 & 0.64 & 0.35 & 0.42 \\ 
\multicolumn{1}{l|}{\cellcolor[HTML]{EFEFEF}16} & 0.24 & 0.22 & 0.39 & 0.25 & 0.13 & 0.23 & 0.09 & 0.20 & 0.05 & -0.08 & 0.12 & 0.11 & 0.49 & 0.48 & 0.58 & 1.00 & 0.18 & 0.50 & 0.42 & 0.34 \\ 
\multicolumn{1}{l|}{\cellcolor[HTML]{EFEFEF}17} & -0.01 & 0.19 & 0.14 & 0.36 & -0.05 & -0.00 & 0.09 & 0.02 & 0.14 & 0.05 & 0.27 & 0.52 & 0.12 & 0.25 & 0.20 & 0.18 & 1.00 & 0.15 & 0.26 & 0.06 \\ 
\multicolumn{1}{l|}{\cellcolor[HTML]{EFEFEF}18} & 0.52 & 0.11 & 0.24 & 0.25 & -0.02 & 0.23 & 0.27 & 0.25 & 0.34 & 0.13 & 0.19 & 0.03 & 0.39 & 0.69 & 0.64 & 0.50 & 0.15 & 1.00 & 0.28 & 0.61 \\ 
\multicolumn{1}{l|}{\cellcolor[HTML]{EFEFEF}19} & 0.24 & 0.36 & 0.05 & 0.25 & 0.23 & 0.13 & 0.03 & 0.35 & 0.06 & 0.14 & 0.23 & 0.08 & 0.13 & 0.38 & 0.35 & 0.42 & 0.26 & 0.28 & 1.00 & 0.27 \\ 
\multicolumn{1}{l|}{\cellcolor[HTML]{EFEFEF}20} & 0.43 & 0.25 & 0.05 & 0.19 & -0.01 & 0.14 & 0.28 & 0.12 & 0.34 & 0.32 & 0.23 & 0.25 & 0.38 & 0.57 & 0.42 & 0.34 & 0.06 & 0.61 & 0.27 & 1.00 \\ 

\end{tabular}
\caption{Correlation matrix for the example set of bonds.}
\label{corrmatr}
\end{table}

\end{landscape}

\bibliography{biblio.bib}

\end{document}